\documentclass[lettersize,journal]{IEEEtran}
\usepackage{amsmath,amsfonts}
\usepackage{algorithmic}
\usepackage{algorithm}
\usepackage{array}
\usepackage[caption=false,font=normalsize,labelfont=sf,textfont=sf]{subfig}
\usepackage{textcomp}
\usepackage{stfloats}
\usepackage{url}
\usepackage{verbatim}
\usepackage{graphicx}
\usepackage{cite}
\usepackage{hyperref}
\usepackage{rotating}
\usepackage{adjustbox}
\usepackage{booktabs}

\begin{document}

\title{Transformer for Multitemporal Hyperspectral Image Unmixing}

\author{Hang Li, Qiankun Dong, Xueshuo Xie, Xia Xu*, Tao Li*,\\
 Zhenwei Shi, \IEEEmembership{Senior Member,~IEEE}

\thanks{This work was supported by the National Natural Science Foundation of China under the Grant 62125102, the Natural Science Foundation of Tianjin under the Grant 23JCQNJC00010 and Natural Science Foundation of Tianjin under Grant 23JCQNJC01050. \textit{(*Corresponding author: Xia Xu and Tao Li.)}}
\thanks{Hang Li, Qiankun Dong and Tao Li are with the College of Computer Science, Nankai University, Tianjin 300071, China (e-mail: lihang\_nk@mail.nankai.edu.cn; qiankund@nankai.edu.cn; litao@nankai.edu.cn).}
\thanks{Xueshuo Xie is with the Haihe Lab of ITAI Street, Tianjin, 300350, China (e-mail: xueshuoxie@nankai.edu.cn).}
\thanks{Xia Xu is with the School of Computer Science and Technology, Tiangong University, Tianjin 300387, China (e-mail: xuxia@nankai.edu.cn).}
\thanks{Zhenwei Shi is with the Image Processing Center, School of
Astronautics, Beihang University, Beijing 100191, China (e-mail:
shizhenwei@buaa.edu.cn).}
}

\markboth{submission to TRANSACTIONS ON IMAGE PROCESSING}%
{Shell \MakeLowercase{\textit{et al.}}: A Sample Article Using IEEEtran.cls for IEEE Journals}


\maketitle

\begin{abstract}
Multitemporal hyperspectral image unmixing (MTHU) holds significant importance in monitoring and analyzing the dynamic changes of surface. However, compared to single-temporal unmixing, the multitemporal approach demands comprehensive consideration of information across different phases, rendering it a greater challenge. To address this challenge, we propose the Multitemporal Hyperspectral Image Unmixing Transformer (MUFormer), an end-to-end unsupervised deep learning model.
To effectively perform multitemporal hyperspectral image unmixing, we introduce two key modules: the Global Awareness Module (GAM) and the Change Enhancement Module (CEM). The Global Awareness Module  computes self-attention across all phases, facilitating global weight allocation. On the other hand, the Change Enhancement Module dynamically learns local temporal changes by comparing endmember changes between adjacent phases. The synergy between these modules allows for capturing semantic information regarding endmember and abundance changes, thereby enhancing the effectiveness of multitemporal hyperspectral image unmixing.
We conducted experiments on one real dataset and two synthetic datasets, demonstrating that our model significantly enhances the effect of multitemporal hyperspectral image unmixing.
\end{abstract}


\section{Introduction}
\IEEEPARstart{H}{yperspectral} remote sensing images have a diverse range of applications and offer conseutive spectral information for Earth observation missions. However, the acquired hyperspectral images (HSIs) are often influenced by sensor limitations, atmospheric conditions, illumination variations, and other factors. Consequently, each pixel in the image may encompass mixed spectral information from multiple ground objects. Hyperspectral image unmixing is conducted to extract endmembers and abundances in HSIs, where endmembers represent pure spectra while abundances denote their respective proportions \cite{r22, r93}. According to different types of mixtures, the assumptions of mixture models can be categorized into linear and nonlinear forms \cite{r95, r77}. The linear mixture model (LMM) is more straightforward to implement as it assumes that the spectrum of each observed pixel can be accurately represented by a linear combination of spectral components from multiple ground objects \cite{r46, r94}.

Due to the simplicity and efficiency of LMM, a lot of studies have been conducted utilizing this methodology, which can be broadly categorized into the following five domains, including geometrical, statistical, nonnegative matrix factorization, sparse regression, and deep learning methods \cite{r22, r65}. Geometrical methods usually use the simplex set or positive cone of the data to expand the study \cite{r39, r97, r40}. Statistical methods rely on parameter estimates and probability distributions for unmixing \cite{r20, r53}. Nonnegative matrix factorization methods estimate endmembers and abundances by factorizing the HSI into two low-rank matrices \cite{r82, r57, r87}. Sparse regression methods define unmixing as a linear sparse regression problem, which is usually studied by using a prior spectral library \cite{r84, r83}. Deep learning approaches primarily employ deep neural networks to extract feature from HSI for unmixing \cite{r90, r32, r33}. 

In recent years, deep learning methods have made remarkable achievements in the field of hyperspectral image unmixing \cite{r65, r90, r85}. As a classical neural network architecture, the autoencoder can map high-dimensional spectral data into a low-dimensional latent space \cite{r66}. Autoencoder cascade comprehensively considers noise and prior sparsity for unmixing through a cascade structure \cite{r67}. Adding constraints through optimizing the loss function of the autoencoder is a commonly employed technique in unmixing \cite{r4}. Some studies use the advantage of variational autoencoders to learn data distribution for unmixing \cite{r68, r69}. Other researchers use autoencoders to extract spatial and spectral information of hyperspectral images for unmixing \cite{r32, r33, r70}. Spectral variability has also attracted extensive research \cite{r92, r98}. The deep learning-based method demonstrates significant efficacy in hyperspectral image unmixing, prompting this paper to also explore research utilizing the deep learning approach.

Current research primarily concentrates on single-phase hyperspectral image unmixing \cite{r91, r77}. However, considering the lengthy revisit period of satellites and their capacity to collect remote sensing data spanning extended durations, leveraging long-term series data proves more beneficial for surface change monitoring. Consequently, harnessing long-term series data for multitemporal hyperspectral image unmixing holds significant value. Compared to change detection of two time phases \cite{r88, r89, r96}, multitemporal unmixing can make use of rich time information to explore the feature correlation of different time points and track the change of endmember and abundance. This is crucial for dynamic earth resource detection, land cover analysis, disaster prevention, and more \cite{r34}. Yet, accurately capturing endmember changes and understanding temporal, spatial, and spectral data pose greater challenges in multitemporal hyperspectral image unmixing.

At present, statistical-based studies are being conducted on the unmixing of multitemporal hyperspectral images. A linear mixing model across time points is explored in \cite{r61}. An online unmixing algorithm, solved via stochastic approximation, is introduced in \cite{r17}. The alternating direction multiplier method is applied in \cite{r18}. KalmanEM employs parametric endmember estimation and Bayesian filtering \cite{r19}. A hierarchical Bayesian model optimizes spectral variability and outliers in \cite{r62}. A Bayesian model considering spectral variability addresses endmember reflectance changes in \cite{r20}. In \cite{r63}, a spectral-temporal Bayesian unmixing method incorporates prior information to mitigate noise effects. An unsupervised unmixing algorithm based on a variational recurrent neural network is proposed in \cite{r21}. 

However, the above methods often require precise definition of prior distributions, which becomes difficult when there are significant changes in endmembers over time. The dynamic nature of land cover and surface conditions leads to varying endmember spectra, making it hard to specify accurate prior distributions for all scenarios. Furthermore, the use of Monte Carlo sampling to estimate posterior distributions in statistical methods adds computational complexity, especially in high-dimensional spaces. Generating samples from the posterior distribution involves iterative procedures, which become more computationally demanding with higher data dimensionality.  

The transformer, known for its powerful attention mechanism, has been widely successful across various fields \cite{r34, r7, r76, r80}. In the realm of unmixing tasks, researchers have explored its potential. For instance, a transformer-based model introduced in \cite{r11} enhances spectral and abundance maps by capturing patch correlations. In \cite{r12}, a window-based transformer convolutional autoencoder tackles unmixing. Another approach in \cite{r13} utilizes a double-aware transformer to exploit spatial-spectral relationships. Meanwhile, \cite{r74} proposes a transformer-based generator for spatial-spectral information. Moreover, a U-shaped transformer network \cite{r75} and methods using spatial or spectral attention mechanisms \cite{r71, r72, r73} are employed for hyperspectral image unmixing. However, these methods primarily focus on single-phase unmixing, lacking the capacity to incorporate time information, which leads to suboptimal performance with multitemporal data. Furthermore, adapting to dynamic relationship changes between phases poses a challenge for traditional transformer models.

To address the challenge of effectively modeling multitemporal hyperspectral image unmixing and adaptively processing the dynamic changes between adjacent phases, we propose the \textbf{M}ulti-temporal Hyperspectral Image \textbf{U}nmixing Transormer (\textbf{MUFormer}). MUFormer is an end-to-end model based on transformer, which consists of two main modules: the Global Awareness Module (GAM) and the Change Enhancement Module (CEM). The GAM synthesizes a comprehensive understanding of the hyperspectral image sequence by computing attention weights for spatial, temporal, and spectral dimensions from a global perspective. This allows for the fusion of multi-dimensional information across the entire image sequence. Conversely, the CEM focuses on capturing nuanced changes in endmember abundances between adjacent phases with high granularity. By assigning adaptive weights to different temporal phases, the sensitivity of the model to time dynamics is enhanced. Through the seamless integration of these modules, MUFormer effectively captures rich multi-dimensional information in multitemporal hyperspectral images. It has adaptability to varying time intervals within multitemporal image sequences, thereby facilitating precise unmixing across different temporal contexts. The main contributions of this paper are as follows:

\begin{itemize}
    \item We propose an end-to-end transformer-based multitemporal hyperspectral image unmixing framework \textbf{MUFormer}, which can achieve effective and efficient unmixing of multitemporal hyperspectral images.
    \item We propose a novel Change Enhancement Module to obtain feature information at different scales and highlight fine changes by multi-scale convolution of adjacent temporal hyperspectral images.
    \item We propose a Global Awareness Module to extract and deeply fuse multitemporal hyperspectral image features from a global perspective, which can better use multi-source domain information to promote the unmixing effect.
\end{itemize}

The remaining of this paper is organized as follows: Section \uppercase\expandafter{\romannumeral2} focuses on the definition of the multitemporal hyperspectral image unmixing task and we mainly introduce our proposed model framework as well as the Change Enhancement Module and the Global Awareness Module. In Section \uppercase\expandafter{\romannumeral3}, we will carry out experiments on a real dataset and two simulated datasets, and conduct ablation experimental research on the proposed module. Finally, we summarize the overall work of this paper in Section \uppercase\expandafter{\romannumeral4}.

\section{Methodology}

\subsection{Multitemporal Hyperspectral Image Unmixing}
For multitemporal hyperspectral images, it is represented as $\mathrm{\mathbf{Y}\in \mathbb{R}^{T\times L\times N}}$ , Where $\mathrm{T}$ represents the number of time phases, $\mathrm{L}$ represents the number of bands of the hyperspectral image, and $\mathrm{N}$ represents the total  number of the hyperspectral image pixles, respectively. For linear unmixing of multitemporal hyperspectral images, the formula for time $t$ is as follows. \\
\begin{equation}
\label{MTHU}
\mathbf{Y_{t} = M_{t}A_{t} + \varepsilon_{t}}
\end{equation}
Where $\mathrm{\mathbf{M_{t}}\in{\mathbb{R}^{L \times P}}}$ represents the matrix of endmembers at time $t$, while $\mathrm{\mathbf{A_{t}}\in{\mathbb{R}^{P \times N}}}$ denotes the abundance matrix corresponding to the same timeframe, and $\mathrm{\mathbf{\varepsilon_{t}}\in{\mathbb{R}^{L \times N}}}$ represents the additional random noise added to simulate the real scenario, where $\mathrm{P}$ denotes the number of endmembers. \eqref{MTHU} can also be refined to the pixel-level form, which satisfies the linear unmixing model for every pixel in every time phase. For the hyperspectral image at each time phase, the abundance nonnegativity constraint (ANC) and the abundance sum-to-one constraint (ASC) are satisfied. 
\begin{equation}
\label{ASC-ANC}
\mathbf{A_{t} \ge 0, 1^{T} A_{t} = 1^{T} }
\end{equation}
In the multitemporal hyperspectral image unmixing task, the hyperspectral images in different time phases have spatial-temporal correlation, and the contribution of hyperspectral bands in each time phase to unmixing is not the same, so it is necessary to jointly consider the temporal-spatial-spectral information to design the unmixing model.

\begin{figure*}[t]
    \centering
    \includegraphics[width=1\linewidth]{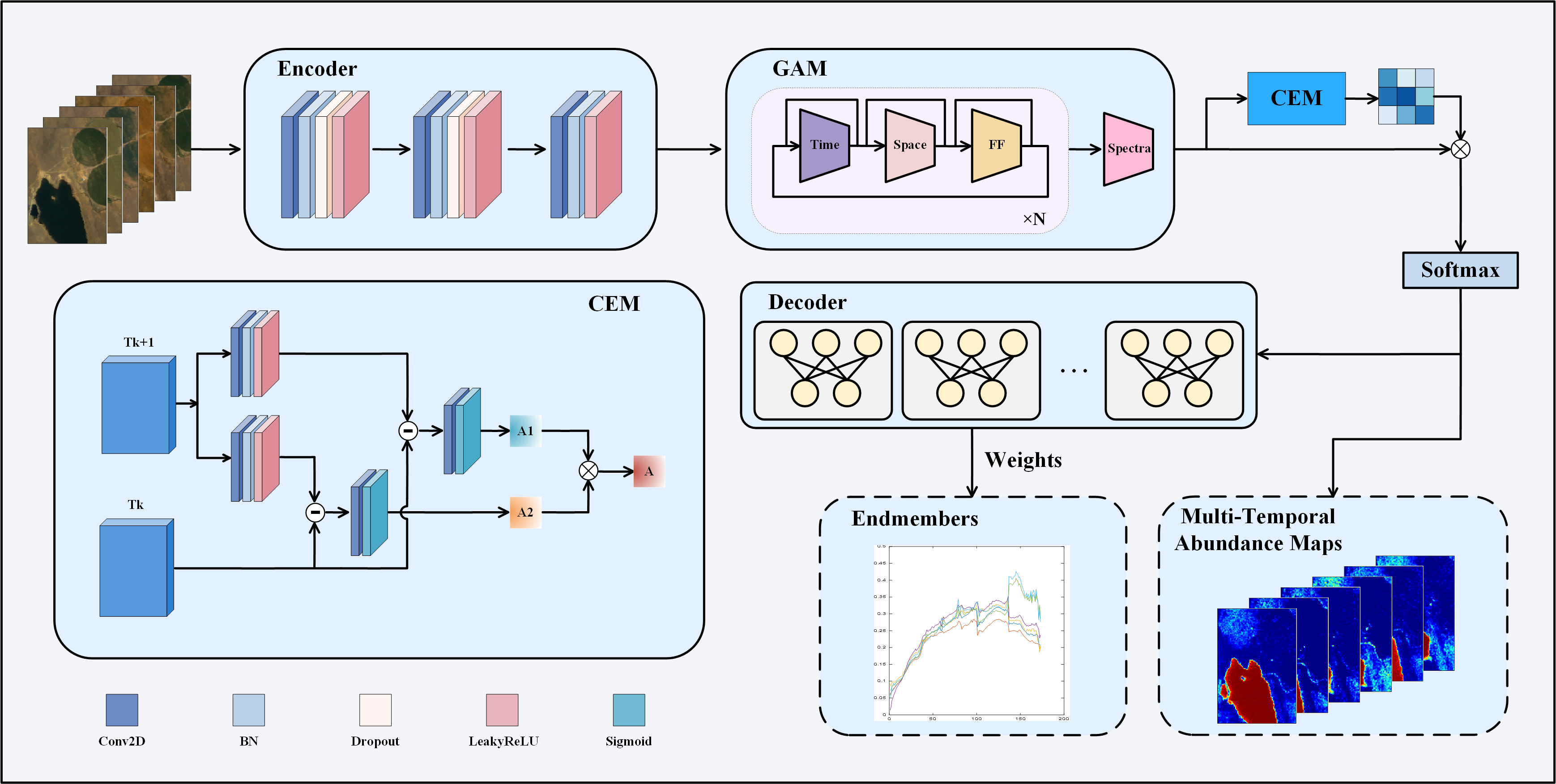}
    \caption{Illustration of the structure of the MUFormer model. Among them, GAM stands for Global Awareness Module, and CBM stands for Change Enhancement Module, as shown in the lower left corner of the illustration.}
    \label{fig:model}  
\end{figure*}

\subsection{Overall structure}
In this section, we introduce the overall structure of the proposed MUFormer. The framework of the MUFormer model is shown in Fig. \ref{fig:model}.

First, we input a sequence of hyperspectral images $\mathbf{Y_{in}}$ to the encoder, where $\mathrm{\mathbf{Y_{in}}\in{\mathbb{R}^{T\times L \times H \times W}}}$, $\mathrm{H}$ and $\mathrm{W}$ represent the height and width of the images, the sequence represents the hyperspectral images acquired at the same location at different time instants. The encoder is mainly composed of convolutional layer, BatchNorm layer, and LeakyReLU layer. In order to prevent the model from overfitting, we add Dropout layer, which makes the neural network more robust and generalization ability \cite{r11}. Compared with transformer, CNN structure has more advantages in extracting local image features, and multitemporal feature maps containing richer semantic information can be obtained through multiple convolution operations \cite{r78, r86, r79}. At this point, the model outputs the multitemporal feature map $\mathbf{Y_{conv}}$, where $\mathrm{\mathbf{Y_{conv}}\in {\mathbb{R}^{T\times C \times H \times W}}}$, $\mathrm{C}$ represents the number of channels retained after convolution processing, and it is a hyperparameter that can be set. Compared with the hundreds of channels of the original hyperspectral image, the channels retained here contain more important band information.

Then, we send the obtained multitemporal feature map into the GAM. With the advantages of transformer in sequence information modeling, we divide the feature map into patch blocks of the same size to calculate attention of each dimension and fuse it. Then, we use the proposed CEM module to refine the changes of adjacent phases, and send the processed feature maps into softmax to obtain the abundance maps of each phase.

In the decoder part, we establish a linear decoder for each phase, incorporating a linear layer. The weight of this linear layer is initialized using the VCA algorithm. Upon completion of the training iteration, the weight of the decoder represents the endmember matrix of the phase. Leveraging the linear decoder, we achieve the realization of the multitemporal linear unmixing model.

\subsection{Global Awareness Module}
In order to calculate temporal-spatial-spectral attention from a global perspective and to deeply integrate information from multiple dimensions, we design the GAM. This module mainly contains temporal attention, spatial attention, spectral attention and forward parts. Feature maps processed by convolutional layers remove redundant bands and contain more critical information. Following the design of ViT \cite{r7}, the feature map sequence is divided into non-overlapping patches $\mathbf{Y_{patch}}$, where $\mathrm{\mathbf{Y_{patch}}\in {\mathbb{R}^{p \times p\times C}}}$, $\mathrm{p}$ represents the side length of the patch and $\mathrm{C}$ represents the number of channels processed by the encoder. We concatenate the $\mathbf{Y_{cls}}$ with the original patch matrix, $\mathbf{Y_{cls}}$ aggregates global features to avoid bias toward specific tokens in the sequence, and then add the position embedding vector $\mathbf{Y_{pos}}$, $\mathrm{\mathbf{Y_{pos}}\in {\mathbb{R}^{D} }}$ represents the temporal-spatial position information of the patch that can be learned. The new matrix $\mathbf{Y^{'}}$ representation is given in \ref{cls}.

\begin{figure*}[t]
    \centering
    \includegraphics[width=1\linewidth]{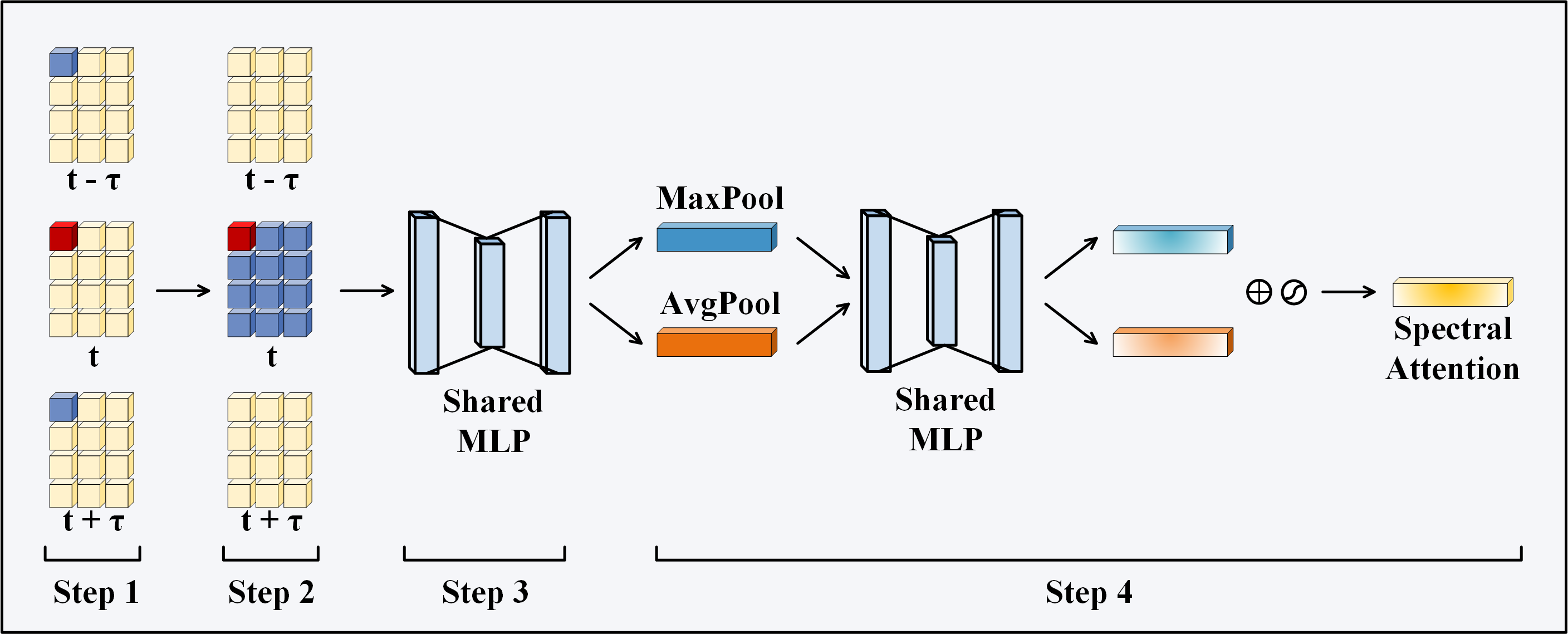}
    \caption{Global Awareness Module, Step1 represents the temporal self-attention calculation process, Step2 represents the spatial self-attention calculation process, Step3 represents the shared MLP module, and Step4 represents the spectral self-attention calculation process. Specifically, in Step1 and Step2, red represents the current query patch, blue represents the patches that participate in the attention computation, and yellow represents the patches that do not participate in the computation.}
    \label{TSSAM}
\end{figure*}

\begin{equation}
\label{cls}            
    \mathbf{Y^{'} = (Y_{cls}  ||  Y_{patch}) + Y_{pos}}
\end{equation}

After obtaining $\mathbf{Y^{'}}$, inspired by the work of \cite{r36}, we treat the hyperspectral image sequence as frames with different time intervals, and then send them to the attention module for processing. Firstly, it is sent to the temporal attention module. In the temporal attention module, $\mathbf{Y}$ will be divided into $\mathbf{Q, K, V}$ three matrices after the linear layer, which represent query, key and value respectively, where $\mathrm{\mathbf{Q}\in {\mathbb{R}^{h_{dim} \times h} }}$, $\mathrm{h_{dim}}$ represents the dimension of each head, $h$ represents the number of heads,the dimensions of $\mathbf{K}$ and $\mathbf{V}$ are the same as $\mathbf{Q}$. After obtaining the above three matrices, the self-attention weight size is calculated through the softmax function, and the calculation formula is as shown in \ref{attn_time}.

\begin{equation}
\label{attn_time}
    \mathbf{\boldsymbol{attn}^{time}_{(l, t)}=\operatorname{Softmax}\left(\frac{\mathbf{q}_{(l, t)}}{\sqrt{D_{h}}} \cdot\left[\mathbf{k}_{(0,0)}\left\{\mathbf{k}_{\left(l, t^{\prime}\right)}\right\}_{\substack{t^{\prime}=1, \ldots, T}}\right]\right)}
\end{equation}

Where $\mathrm{l}$ represents the $\mathrm{lth}$ patch block and $\mathrm{t}$ represents the hyperspectral image at time $\mathrm{t}$. By computing attention in the time dimension, we can assign reasonable weights to patches at the same location but at different moments, this enables the model to identify regions that have undergone changes at different times. By allocating higher attention to patches that exhibit significant changes throughout the time series, the model can concentrate more effectively on these key areas of change. Similarly, for patches at different positions at the same time, we also use the same way to calculate the attention weights of spatial dimensions, as shown in \ref{attn_space}.

\begin{equation}
\label{attn_space}
    \mathbf{\boldsymbol{attn}^{space}_{(l, t)}=\operatorname{Softmax}\left(\frac{\mathbf{q}_{(l, t)}}{\sqrt{D_{h}}} \cdot\left[\mathbf{k}_{(0,0)}\left\{\mathbf{k}_{\left(l^{\prime}, t\right)}\right\}_{\substack{l^{\prime}=1, \ldots, N}}\right]\right)}
\end{equation}

In order to solve the problem of gradient disappearance and gradient explosion in deep networks, we adopt residual connection \cite{r37}. The feature maps processed by temporal self-attention and spatial self-attention are fed into the forward module, which contains an MLP. This module can be set to multiple continuous modules, in order to prevent overfitting of the model here the model depth is set to 2. Different channels in hyperspectral images may correspond to different ground object components or features, and some channels may be affected by noise, atmospheric interference and other factors, resulting in low information quality. At the same time, the contribution of different channels to the unmixing task is not consistent \cite{r38, r64}. Combined with the above reasons, we send it to the spectral self-attention calculation module. In the spectral self-attention module, we respectively use Average pooling and Max pooling to process the channels, and then generate the channel attention vector through the full connection operation, which can adaptively adjust the weight of each channel to highlight important channel information. The ability to focus on channels with higher unmixing contributions is enhanced to improve the unmixing accuracy.The spectral self-attention mechanism is calculated as shown in \ref{cbam}. $\mathbf{Y_{map}}$ represents the feature map outputted by FF, and $\sigma$ represents the sigmoid function. The schematic diagram of GAM is shown in Fig. \ref{TSSAM}.

\begin{equation}
    \label{cbam}
    \mathrm{\mathbf{Y^{\prime \prime }} = \sigma (MLP(Avgpool(\mathbf{Y_{map}} )+ Maxpool(\mathbf{Y_{map}})))}
\end{equation}

Through the GAM, we dynamically assign weights to the significance of the time, space, and spectral dimensions, enabling the more precise capture of complex features within image data. This holistic attention mechanism aids models in more effectively comprehending the temporal evolution of objects or phenomena, alongside their unique spatial distributions and spectral characteristics. Concurrently, the robustness of the model is enhanced as it emphasizes crucial features while diminishing the impact of irrelevant or disruptive information. This enhancement makes our model highly skilled in dealing with the complex and fluctuating environmental conditions prevalent in hyperspectral images.

\subsection{Change Enhancement Module}
Through the above GAM, we can obtain the temporal-spatial-spectral feature information from a global perspective. However, the acquisition time interval of hyperspectral images in different time phases is not certain. Therefore, how to adaptively deal with the feature changes of hyperspectral images between adjacent time phases is also an important issue. In order to solve this problem, we propose the CEM from the feature-level perspective, and its structure diagram is shown in the bottom left corner of Fig. \ref{fig:model}. The CEM input is the feature map of two adjacent time phases after GAM processing, which size is $\mathrm{{\mathbb{R} }^{C\times H\times W}}$. At this time, the feature map has gone through the weight allocation of global attention. In order to accurately capture the difference between adjacent temporal hyperspectral images from the feature level better, we propose to perform multi-scale convolution on the feature map at time $\mathrm{T_{k+1}}$, where the convolution kernel size is $3\times3$ and $7\times7$, respectively. Then the feature maps after convolution are subtracted from the feature graphs at $\mathrm{T_{k}}$ time. The subtracted feature maps are fed into the convolution layer and processed by the $Sigmoid$ function, where the kernel size is set to $1\times1$. The weight vector $\mathbf{A}$ is obtained by multiplying the obtained weights $\mathbf{A1}$ and $\mathbf{A2}$, and the final weight vector is obtained by multiplying $\mathbf{A}$ by $\alpha$. We add the obtained weight vector to $\mathbf{Y_{cls}}$, $\mathbf{Y_{cls}}$ tags can provide a high-level representation of the entire sequence to facilitate learning and inference by the model. It allows the model to extract important global information from the input sequence rather than just focusing on local segments \cite{r45}. We modify $\mathbf{Y_{cls}}$ by combining the output of CEM, so that the model not only allocates global information weights through the attention mechanism, but also uses convolutional neural network to reasonably deal with the change regions of adjacent time phases in the local information extraction. By this way of ``coarse tuning'' (GAM) plus ``fine tuning'' (CEM), we can reasonably simulate the changes of multitemporal features, so that our model can adaptively unmix multitemporal hyperspectral images.

\begin{equation}
    \label{A1}
    \mathrm{\mathbf{A1}=\sigma (f_{1\times 1}(f_{3\times 3}(\mathbf{T_{k+1}})-\mathbf{T_{k}})), 1\le k\le T-1}
\end{equation}

\begin{equation}
    \label{A2}
    \mathrm{\mathbf{A2}=\sigma (f_{1\times 1}(f_{7\times 7}(\mathbf{T_{k+1})-T_{k}})), 1\le k\le T-1}
\end{equation}

\begin{equation}
    \label{A2-equ}
    \mathbf{Y_{cls} = (1 + \alpha \times (A1 \times A2))\times Y_{cls}}
\end{equation}
Where $\mathrm{f_{3\times3}}$ and $\mathrm{f_{7\times7}}$ represent convolution kernel sizes of 3 and 7, respectively, followed by BN layer and LeakyReLU layer. $\mathrm{f_{1\times1}}$ represents a convolutional layer with a kernel size of 1 and $\sigma$ represents the $Sigmoid$ function, where $\alpha$ is a hyperparameter that can be set empirically. We perform multi-scale convolution on the feature map, which can capture features at different scales. The smaller $3\times3$ kernels help to capture local details and edge features, while the larger $7\times7$ kernels are more suitable to capture larger contextual information. By using convolution kernels at both scales, a more comprehensive multi-scale feature representation can be obtained. In the process of feature fusion, the information loss that may exist in a single scale can be reduced and the expression ability of the model can be improved. Moreover, the large-scale convolution kernel may be more effective for removing some noise and unimportant detail information in the image, while the small-scale convolution kernel helps to retain important local features, which also suppresses noise to a certain extent. The ablation study on multi-scale CEM can be seen in Section \uppercase\expandafter{\romannumeral4}.

\subsection{Loss Function}
In the design of loss function, we choose the weighted sum of multiple loss functions as the total loss function to better train our proposed model. Reconstruction loss and Spectral Angle Distance (SAD) loss are commonly used loss functions in unmixing tasks. At the same time, it was demonstrated in \cite{r40} that the data simplex loss plays an important role in endmember estimation, the endmembers can be further constrained by adding the data simplex loss. The total training loss function is shown in \ref{loss}, including $\mathrm{L_{RE}}$, $\mathrm{L_{SAD}}$ and $\mathrm{L_{E}}$, where $\beta$, $\gamma$ and $\lambda$ are artificially set hyperparameters to balance each loss function. We adopt Adam algorithm for optimization and set scheduler to dynamically adjust the learning rate.

\begin{equation}
    \label{re_loss}
    \mathrm{L_{\mathrm{RE}}(\mathbf{Y}, \hat{\mathbf{Y}})=\left (\frac{1}{H \cdot W \cdot T} \sum_{i=1}^{H} \sum_{j=1}^{W}\sum _{t=1}^{T}\left(\hat{\mathbf{Y}}_{\mathbf{i j t}}-\mathbf{Y}_{\mathbf{i j t}}\right)^{2}\right )^{\frac{1}{2} }}
\end{equation}

\begin{equation}
    \mathrm{L_{\mathrm{SAD}}(\mathbf{Y}, \hat{\mathbf{Y}})={\frac{1}{R}\frac{1}{T}  \sum_{i=1}^{R}\sum_{t=1}^{T}  \arccos \left(\frac{\left\langle\mathbf{Y}_{it}, \hat{\mathbf{Y}}_{it}\right\rangle}{|| \mathbf{Y}_{it} ||_{2} ||  \hat{\mathbf{Y}}_{it} ||_{2} }\right)}}
\end{equation}

\begin{equation}
    \mathrm{L_{E}=\sum_{t=1}^{T} \mathbf{\left \| E_{t}  -m_{t} 1_{r}^{T} \right \| _{2}^{F}}}
\end{equation}

\begin{equation}
\label{loss}                
    \mathrm{L=\beta L_{\mathrm{RE}} + \gamma L_{\mathrm{SAD}} + \lambda L_{\mathrm{E}}}
\end{equation}

\begin{figure}
    \centering
    \includegraphics[width=1\linewidth, height=0.32\linewidth]{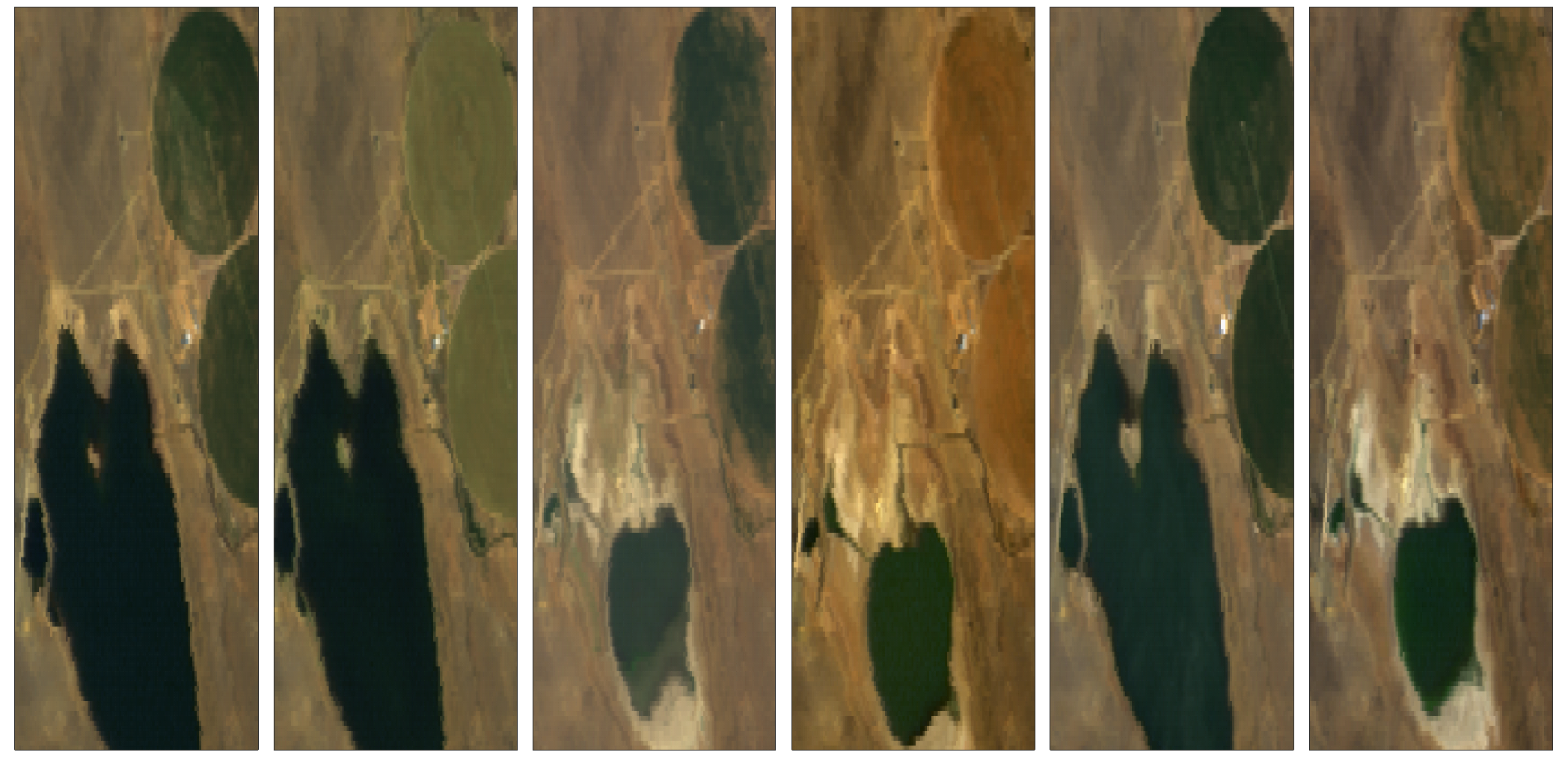}
    \caption{Lake Tahoe hyperspectral image sequence, acquisition time from left to right are 04/10/2014, 06/02/2014, 09/19/2014, 11/17/2014, 04/29/2015, 10/13/2015, respectively.}
    \label{hsi_lake}
\end{figure}

\begin{figure*}[!htbp]
    \centering
    \subfloat[]{\includegraphics[width=2.2in, height=2.8in]{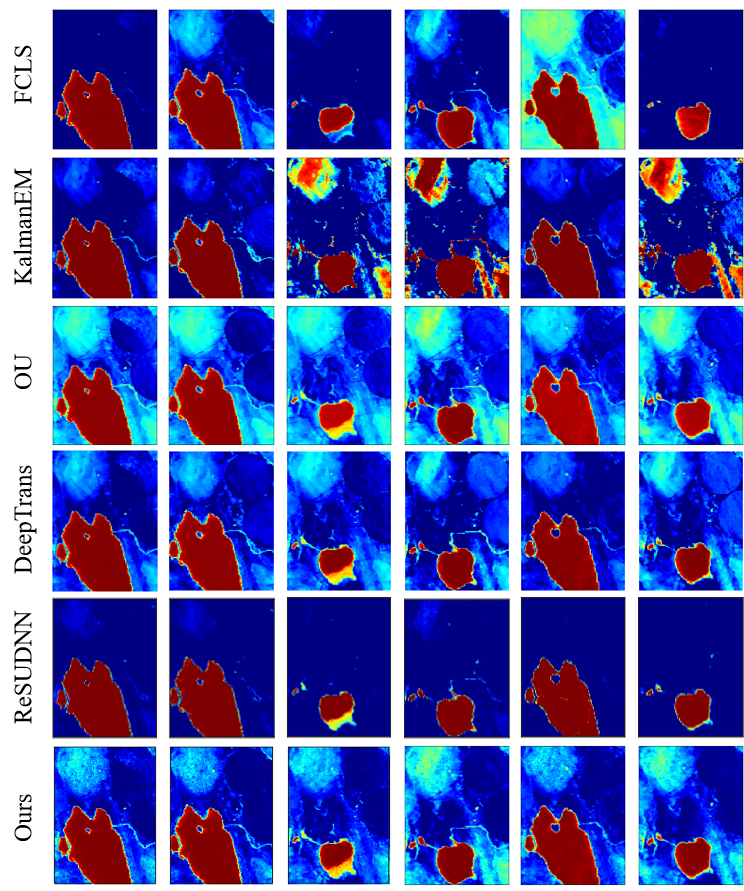}}%
    \label{lake}
    \hfil
    \subfloat[]{\includegraphics[width=2.2in, height=2.8in]{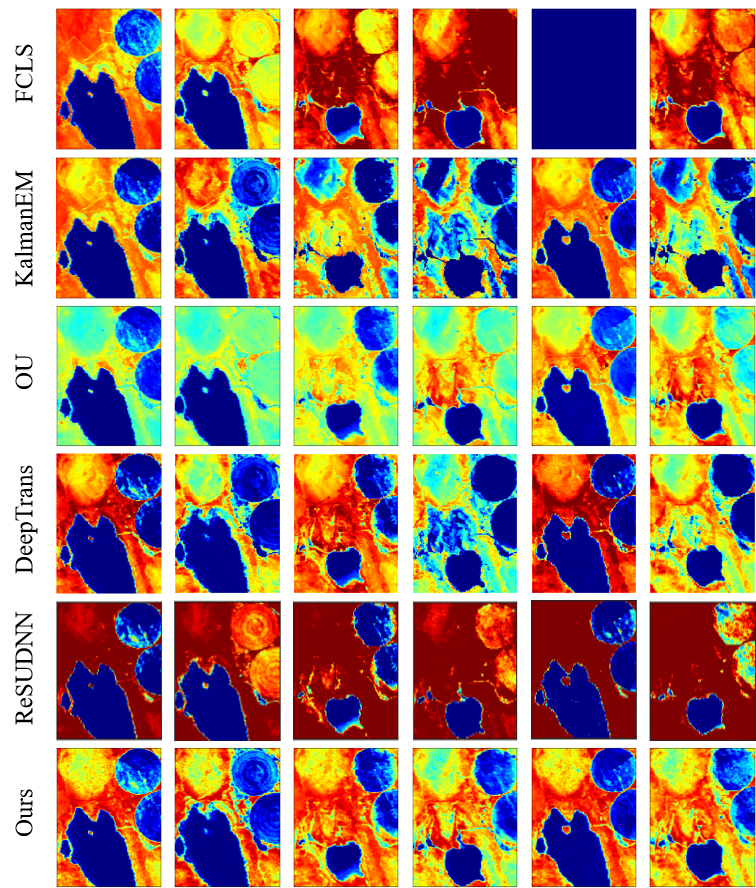}}%
    \label{soil}
    \hfil
    \subfloat[]{\includegraphics[width=2.2in, height=2.8in]{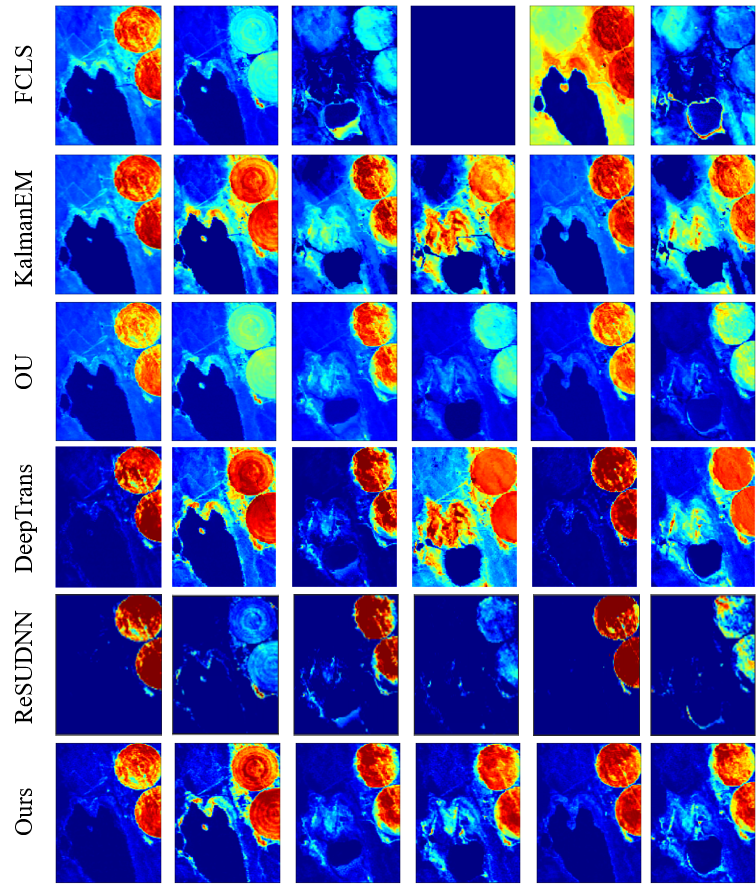}}%
    \label{veg}
    \caption{Multitemporal abundance map of Lake Tahoe HIs, with water, soil, and vegetation endmembers from left to right.}
    \label{Lake-abu}
\end{figure*}

\begin{figure*}
    \centering
    \includegraphics[width=1\linewidth]{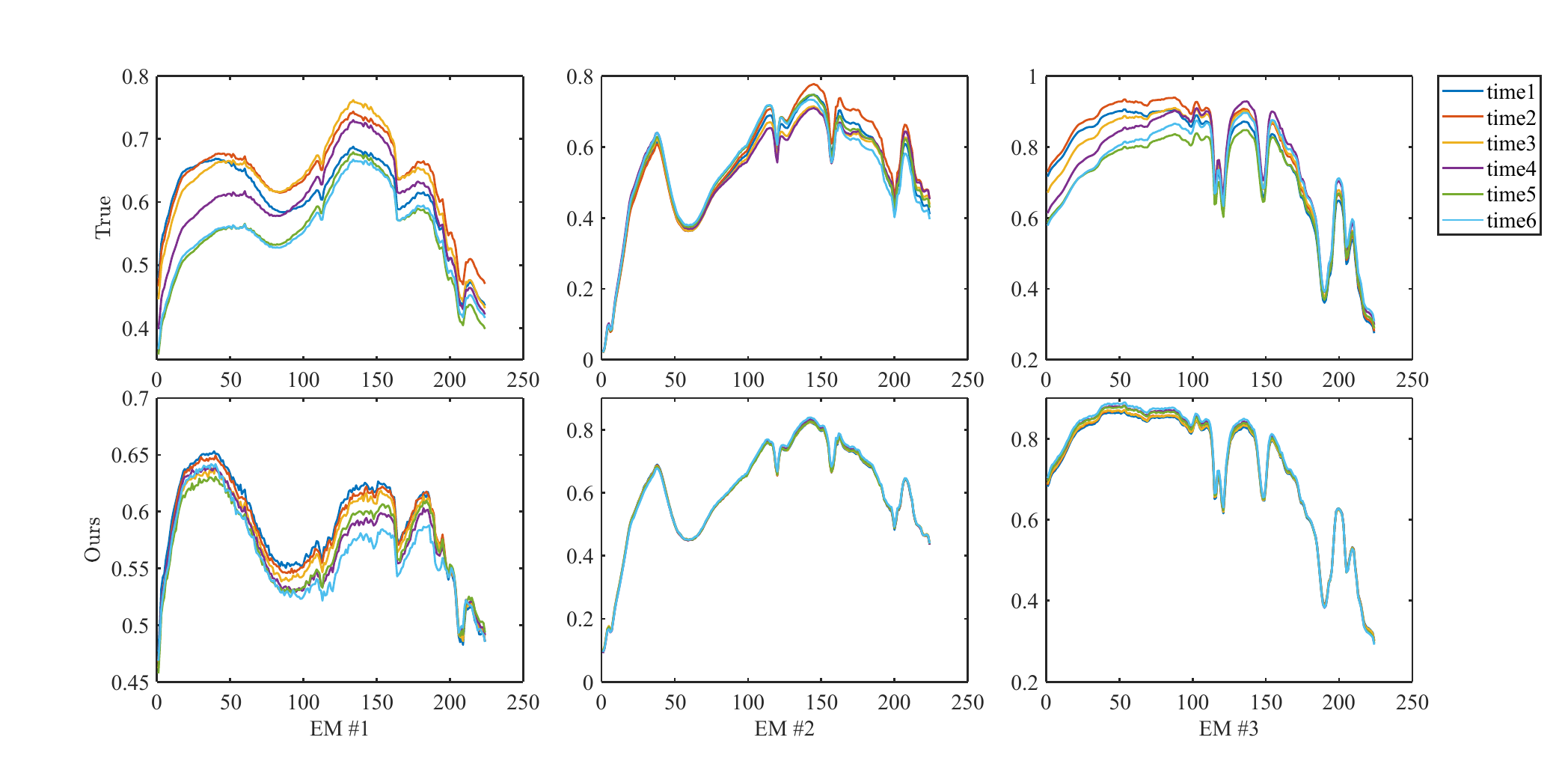}
    \caption{The endmember estimation results for synthetic data 1, where the first row represents the true endmember results and the second row is the result of our model.}
    \label{end-syn1}
\end{figure*}

\begin{figure*}
    \centering
    \includegraphics[width=1\linewidth]{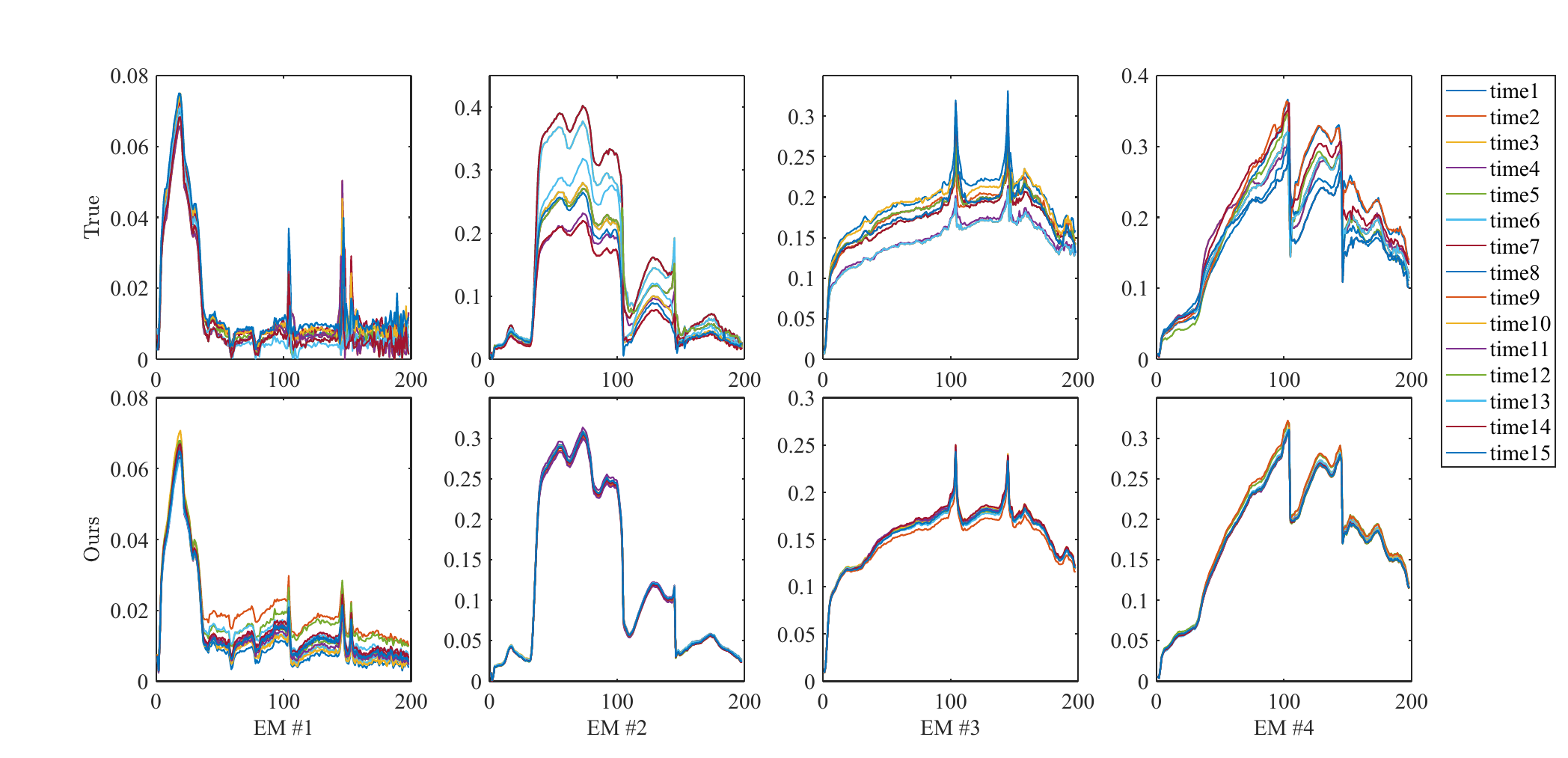}
    \caption{The endmember estimation results for synthetic data 2, where the first row represents the true endmember results and the second row is the result of our model.}
    \label{end-syn2}
\end{figure*}

\section{Experimental Results}
To verify the effectiveness of MUFormer, we conducted experiments on one real dataset and two synthetic datasets. We compare our method with fully constrained least squares (FCLS), online unmixing (OU) \cite{r17}, KalmanEM based on Kalman filter and maximum expectation strategy \cite{r19}, DeepTrans based on transformer for single phase unmixing \cite{r11}, and ReSUDNN based on variational RNN for dynamic unmixing \cite{r21}. For the single-phase unmixing algorithm, we unmix the hyperspectral images of each phase, and then merge each phase to get multitemporal results. The initialization endmember of the above method is obtained by the VCA algorithm.

In order to accurately measure the results of the experiment, we choose two main evaluation metrics. The first one is the normalized root mean square error (NRMSE), which we calculate for the abundance map, endmembers and reconstructed hyperspectral images. The second is the spectral angle mapper (SAM), whose SAM we computed for the endmember. Where $\mathbf{a}$ represents the true abundance value of the $\mathrm{n}$th pixel at the $\mathrm{t}$th time, $\mathbf{\hat{a}_{n,t}}$ represents the estimated abundance value, $\mathbf{\hat{M}_{n,t}}$ represents the estimated endmember value, $\mathbf{m_{n,t,p}}$ represents the $\mathrm{p}$th endmember value in the $\mathrm{n}$th pixel at the $\mathrm{t}$th time, similarly, $\mathbf{\hat{m}_{n,t,p}}$ represents its estimated value. 

For MTHU task, we don't consume too much computing power, we use i7-11800H CPU and RTX 3060 GPU to complete the task efficiently. For these sets of experiments, we set the number of transformer blocks to 2, and too deep networks will cause overfitting or performance degradation of the model. We set the number of attention heads to 8, and the number of epochs during the model training to 1000, so that the model is fully trained.

\begin{equation}
    \mathrm{NRMSE_{\mathbf{A}}=(\frac{1}{T} {\textstyle \sum_{t=1}^{T}} {\textstyle \sum_{n=1}^{N}} \mathbf{\frac{\left \| a_{n,t}-\hat{a}_{n,t}  \right \| ^{2}}{\left \| a_{t} \right \|^{2} }}   )^{\frac{1}{2} }}
\end{equation}

\begin{equation}
    \mathrm{NRMSE_{\mathbf{M}}=(\frac{1}{NT} {\textstyle \sum_{t=1}^{T}} {\textstyle \sum_{n=1}^{N}} \mathbf{ \frac{\left \| M_{n,t}-\hat{M}_{n,t}  \right \|_{F} ^{2}}{\left \| M_{n,t} \right \|_{F}^{2}} }   )^{\frac{1}{2} }}
\end{equation}

\begin{equation}
    \mathrm{NRMSE_{\mathbf{Y}}=(\frac{1}{T} {\textstyle \sum_{t=1}^{T}} {\textstyle \sum_{n=1}^{N}} \mathbf{\frac{\left \| y_{n,t}-\hat{M}_{n,t}\hat{a}_{t}   \right \|^{2}}{\left \| y_{n,t} \right \|^{2} } }   )^{\frac{1}{2} }}
\end{equation}

\begin{equation}
    \mathrm{SAM_{\mathbf{M}}=\frac{1}{TNP} {\textstyle \sum_{t=1}^{T}} {\textstyle \sum_{n=1}^{N}} {\textstyle \sum_{p=1}^{P}}\arccos \mathbf{(\frac{m_{n,t,p}^{\top } \hat{m}_{n,t,p} } {\left \| m_{n,t,p} \right \| \left \| \hat{m}_{n,t,p} \right \| }  )}  }            
\end{equation}

\subsection{Dataset Description}
In order to accurately test the effect of our proposed model, we expand the test on three datasets, including one real dataset and two synthetic datasets, which will be described in the following expansion.
\begin{enumerate}
    \item Lake Tahoe: The dataset was acquired by the airborne Visible Infrared Imaging Spectrometer (AVIRIS) between 2014 and 2015 and contains hyperspectral images of six time phases. The size of the hyperspectral image in each time phase is $150\times110$ pixels, and there are 173 bands after removing the water absorption band. The image contains three types of endmembers: water, soil, and vegetation, each of which produces distinct changes over time \cite{r17}. This sequence of images is shown in Fig. \ref{hsi_lake}.
    \item Synthetic data1: The dataset contains six temporal hyperspectral synthetic images, each of size $50\times50$ pixels, containing three endmembers, and three features of $\mathrm{L=224}$ bands randomly sampled from the USGS library as the reference endmember matrix. To model the endmember variability, the endmembers of each pixel are obtained by multiplying the reference signatures with a piecewise linear random scaling factor of the amplitude interval [0.85,1.15]. Local pixel mutation is added to $t\in \left \{ 2,3,4,5 \right \}$  to better match the endmember changes of real scene, To simulate realistic scenarios, Gaussian noise with a SNR of 30dB was added \cite{r21}.
    \item Synthetic data2: The second synthetic dataset contains 15 temporal hyperspectral images, each $50\times50$ pixels in size. In order to introduce realistic spectral variability, the endmember features of each pixel and phase are randomly selected from the artificially extracted pure pixels of water, vegetation, soil, and road in Jasper Ridge HI, which contains 198 bands. To simulate realistic scenarios, Gaussian noise with a SNR of 30dB was added \cite{r21}.
\end{enumerate}

\subsection{Lake Tahoe Results}
\begin{figure}
    \centering
    \includegraphics[width=1\linewidth]{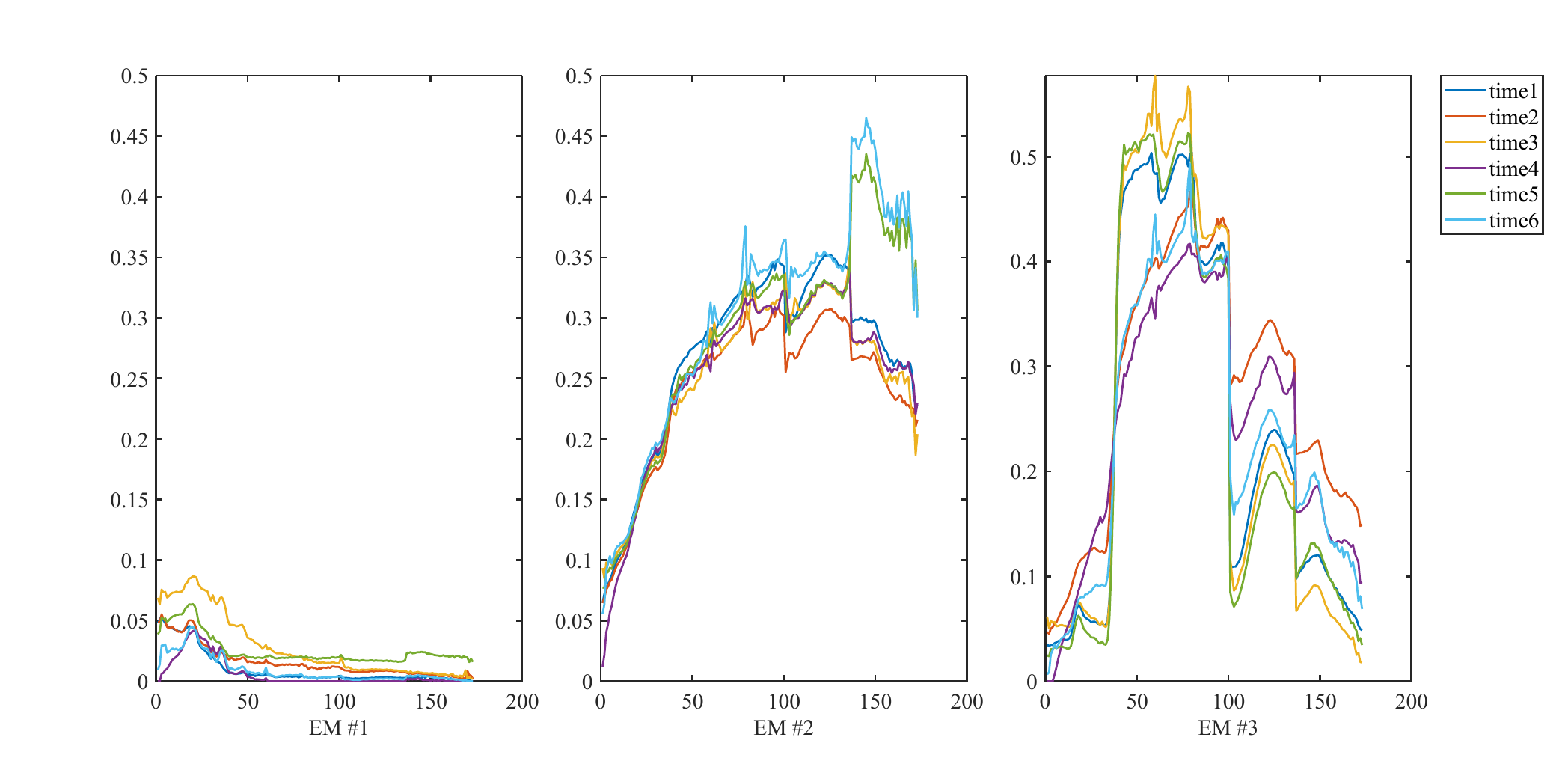}
    \caption{Multitemporal endmember estimation results for the Lake Tahoe dataset.}
    \label{lake-end}
\end{figure}

Due to the lack of real comparison results for the Lake Tahoe dataset, we only present its abundance map qualitatively, and the results are shown in Fig. \ref{Lake-abu}. We alse show the results of multitemporal endmember estimation results in Fig. \ref{lake-end}. It can be seen that the MUFormer proposed by us achieves better results in the unmixing of real multitemporal hyperspectral datasets. We can see from the results that FCLS has the worst performance, and the endmember is missing in multiple phases. There is also confusion between different endmembers. KalmanEM and OU have similar effects, but there are large errors in the abundance estimation of water and soil endmembers. Kalman filter has shortcomings in a large range of abundance changes and there are a lot of artifacts, which is because the Kalman filter sets the abundance to be continuous. When there is a large change in abundance, the effect will be affected. At the same time, in order to form a reasonable comparison, we add the DeepTrans model, which has superior performance on single-temporal hyperspectral image unmixing, as one of the comparison models. We can see that DeepTrans performs well in some individual temporal phases, but it does not consider the temporal information of multitemporal hyperspectral images, resulting in more erroneous unmixing regions. Finally, by comparing our proposed method with ReSUDNN, it can be seen that on the whole, both of them have achieved good results, and they are meticulous in the edge processing of endmembers. However, ReSUDNN has the phenomenon of misjudgment of endmembers in some phases, especially in the estimation of soil and vegetation.  

\begin{figure}[!htbp]
    \centering
    \includegraphics[width=1\linewidth]{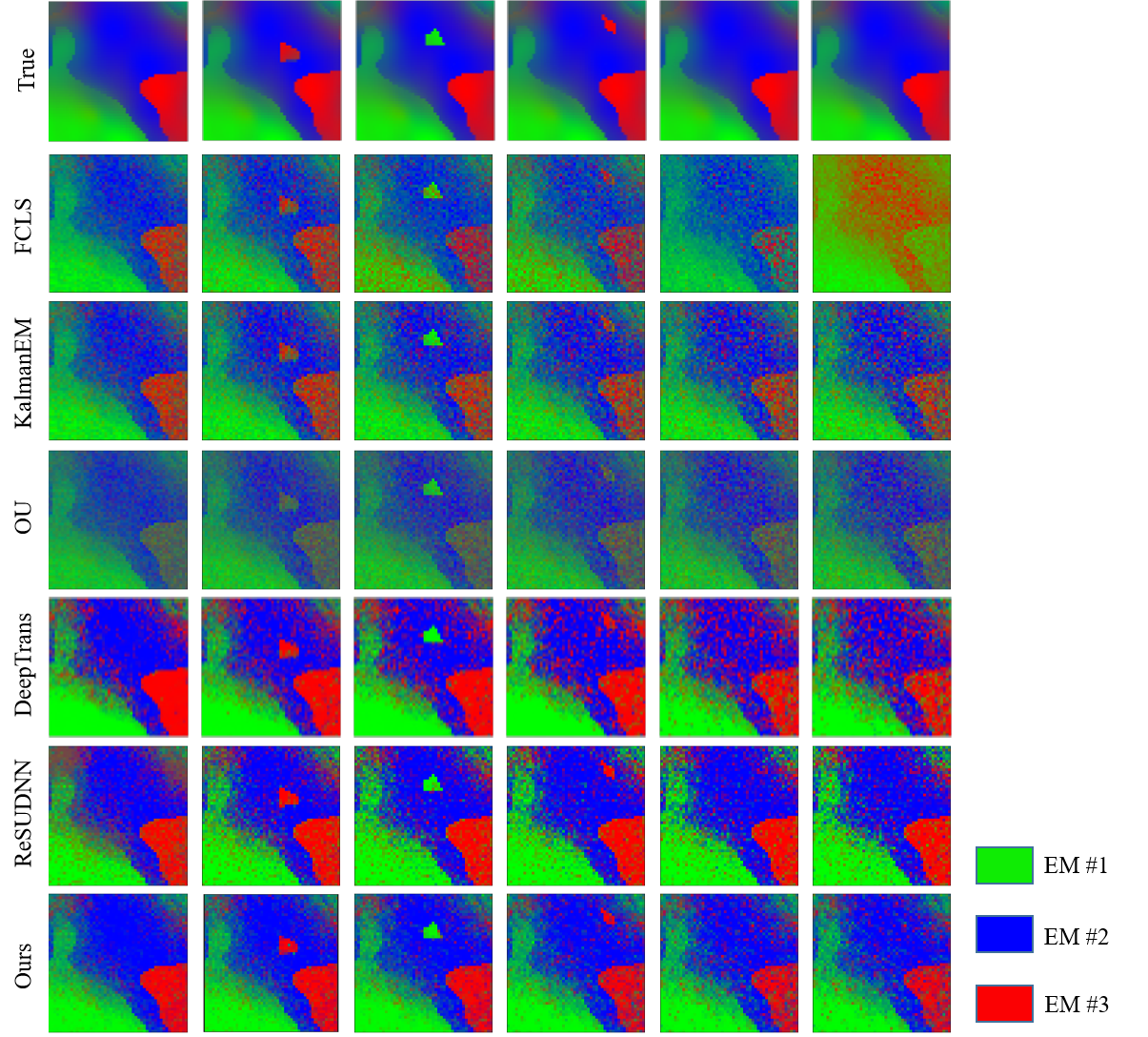}
    \caption{Abundance map estimation results for synthetic dataset 1, where green, blue, and red represent the three endmembers, respectively.}
    \label{syn-abu}
\end{figure}

\begin{figure*}
    \centering
    \includegraphics[width=1\linewidth, height=0.66\linewidth]{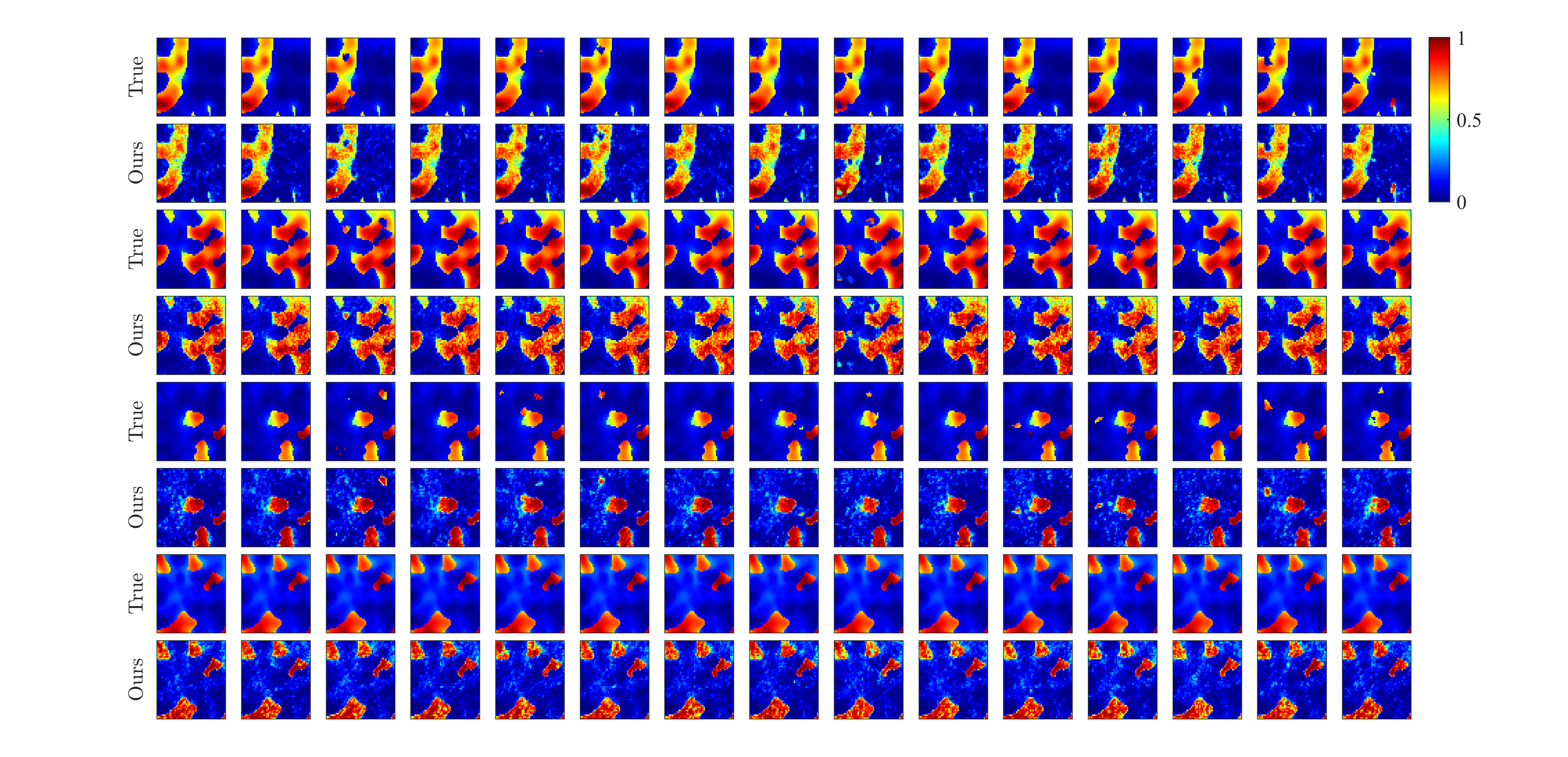}
    \caption{The results of abundance estimation of the synthesized dataset 2. Comparison of the true abundance maps of four endmembers with the estimated abundance maps of the model, where each two-row abundance maps represent a set of comparison results for the same endmember.}
    \label{abu-syn2}
\end{figure*}

\subsection{Synthetic Datasets Results}
Due to the lack of ground truth in Lake Tahoe, we can only qualitatively compare the abundance map results. In order to compare their unmixing accuracy more precisely, we also tested them on two synthetic datasets.  We performed a quantitative comparison on synthetic datasets, the results of which are shown in Table \ref{syn1}. The results of the abundance maps for each model are shown in Fig. \ref{syn-abu}. Since the synthetic dataset 2 contains 15 phases, we only compare the MUFormer estimated abundance plot with the real one, taking into account space constraints. The comparison results are shown in Fig. \ref{abu-syn2}, and the quantitative results are shown in Table \ref{syn2}.

\begin{table}
    \centering
    \caption{Quantitative comparison results in synthetic dataset 1, where the optimal result is shown in bold.}
    \resizebox{\linewidth}{!}{
    \begin{tabular}{cccccc}
        \toprule
         & $\mathrm{NRMSE_{A}}$ & $\mathrm{NRMSE_{M}}$ & $\mathrm{SAM_{M}}$ & $\mathrm{NRMSE_{Y}}$ & $\mathrm{Time}$ \\
         \midrule
       FCLS  & 0.537 & - & - & 0.086 & \textbf{2.7}\\
       OU & 0.434 & 0.342 & 0.260 & \textbf{0.059} & 24.9 \\
       KalmanEM & 0.356 & 0.124 & 0.076 & 0.061 & 2422.8\\
       DeepTrans& 0.592 & 0.381 & 0.286 & 0.133 & -\\
       ReSUDNN  & 0.318 & 0.117 & 0.075 & 0.089 & 479.0\\
       Ours  & \textbf{0.255} & \textbf{0.110} & \textbf{0.059} & 0.079 & 132.2\\
         \bottomrule
    \end{tabular}
    }
    \label{syn1}
\end{table}

\begin{table}
    \centering
    \caption{Quantitative comparison results in synthetic dataset 2, where the optimal result is shown in bold.}
    \resizebox{\linewidth}{!}{
    \begin{tabular}{cccccc}
        \toprule
         & $\mathrm{NRMSE_{A}}$ & $\mathrm{NRMSE_{A}}$ & $\mathrm{SAM_{M}}$ & $\mathrm{NRMSE_{Y}}$ & $\mathrm{Time}$ \\
         \midrule
       FCLS  & 0.500 & - & - & 0.122 & \textbf{7.3}\\
       OU & 0.335 & 0.256 & 0.120 & \textbf{0.055} & 60.6 \\
       KalmanEM & 0.659 & 12.222 & 0.496 & 0.108 & 5937.4\\
       DeepTrans& 0.723 & 0.873 & 0.665 & 0.384 & -\\
       ReSUDNN  & 0.294 & 0.203 & 0.289 & 0.160 & 1231.3\\
       Ours  & \textbf{0.185} & \textbf{0.162} & \textbf{0.100} & 0.153 & 223.0\\
         \bottomrule
    \end{tabular}
    }
    \label{syn2}
\end{table}

\textit{1) Discussion:} From Fig. \ref{syn-abu} and \ref{abu-syn2}, we can see that each model is able to capture the abundance change from time t=2 to time t=6, but all are affected by noise to varying degrees. FCLS has a large estimation error at time t=6, OU and KalmanEM are seriously polluted by noise, and DeepTrans, as a single-temporal unmixing model, performs better at the first two moments, but it does not consider the time information, so the effect is poor for the latter moments, and there is a case of wrong estimation. As a whole, ReSUDNN and our method are closer to the true abundance map. We can see that for long time sequence hyperspectral images, our model shows excellent results, which is because transformer has a good effect on long sequence information processing. At the same time, we also combine the GAM to process the abundance transformation information of adjacent time phases, which further enhances the effect of our model in processing multitemporal hyperspectral images. In Table \ref{syn1} and \ref{syn2} we compare the quantitative results of the individual algorithms, it can be seen that our proposed MUFormer achieves SOTA results in three indicators of $\mathbf{NRMSE_{A}}$, $\mathbf{NRMSE_{M}}$ and $\mathbf{SAM_{M}}$. Although OU has achieved good results on $\mathbf{NRMSE_{Y}}$, the reconstruction loss cannot explain the effect of abundance and endmember estimation, so we see that his results of $\mathbf{NRMSE_{A}}$, $\mathbf{NRMSE_{M}}$ and $\mathbf{SAM_{M}}$ are poor, and the abundance map obtained is also affected by a lot of noise. and at the same time, the running speed is several times higher than that of KalmanEM and ReSUDNN. It is worth mentioning that synthetic dataset 2 contains 15 temporal phases, which is more challenging than synthetic dataset 1. We can see that our method has a substantial improvement, which also highlights the potential of our model in processing long sequences of multitemporal hyperspectral images. In Fig. \ref{end-syn1} and Fig. \ref{end-syn2}, we show the results of endmember estimation of synthetic dataset 1 and 2. It can be seen that our estimation results are very close to the real endmember, but we still need to further enhance the perception ability of the model for the change of phase at different times of the same endmember.

\subsection{Ablation study}
To fully verify the role played by our proposed CEM and GAM in the model, we launched an ablation study, the results of which are shown in Table \ref{ablation-1}. From Table \ref{ablation-1}, we can see that the model performance is improved to varying degrees when the CEM module and GAM are added separately, and the model effect is the best when the two modules are added at the same time. This is because through the GAM, we extract the spatial, global time and spectral information features of the multitemporal hyperspectral image. Combined with the CEM module to capture the changes between adjacent time phases, it is more conducive to our model to deal with multitemporal tasks. At the same time, as mentioned above, we processed images at different scales in CEM. In order to explore the influence of different scale processing on the unmixing effect, we carried out ablation experiments on the size of the convolution kernel, and the experimental results are shown in Table \ref{ablation-2}. From Table \ref{ablation-2}, we can see that the addition of convolution modules of different scales to $cls$ has different degrees of improvement, and the fusion of features at different scales can improve the robustness of the model. We compare the two fusion methods of addition and multiplication, and it can be seen that the multiplication effect of A1 and A2 is more obvious. We believe that multiplying them enables the model to pay more attention to the relationship between features at different scales, which helps to capture the information in the image more comprehensively, especially in the presence of multiple scale structures. Moreover, the correlation between two features can be emphasized by multiplication, and the weighted attention to different regions can also be realized, making the model more flexible to learn different degrees of attention to different locations.

\begin{table}
    \centering
    \caption{Results of ablation studies for each module of the model.}
    \resizebox{\linewidth}{!}{
    \begin{tabular}{cccccc}
        \toprule
        Settings & $\mathrm{NRMSE_{A}}$ & $\mathrm{NRMSE_{M}}$ & $\mathrm{SAM_{M}}$ & $\mathrm{NRMSE_{Y}}$ \\
         \midrule
       Baseline  & 0.362 & 0.111 & 0.061 & 0.083\\
       + CEM  & 0.274 & 0.112 & 0.060 & \textbf{0.070} \\
       + GAM & 0.293 & 0.115 & 0.064 & 0.086\\
       MUFormer& \textbf{0.255} & \textbf{0.110} & \textbf{0.059} & 0.079\\
         \bottomrule
    \end{tabular}
    }
    \label{ablation-1}
\end{table}

\begin{table}
    \centering
    \caption{Results of ablation studies for each module of the model.}
    \resizebox{\linewidth}{!}{
    \begin{tabular}{cccccc}
        \toprule
        Settings & $\mathrm{NRMSE_{A}}$ & $\mathrm{NRMSE_{M}}$ & $\mathrm{SAM_{M}}$ & $\mathrm{NRMSE_{Y}}$ \\
         \midrule
       Baseline  & 0.293 & 0.115 & 0.064 & 0.086\\
       A1  & 0.260 & 0.113 & 0.060 & \textbf{0.077} \\
       A2  & 0.262 & 0.111 & 0.060 & 0.079\\
       A1 + A2 & 0.267 & 0.111 & 0.061 & 0.085\\
       $A1 \times A2$ & \textbf{0.255} & \textbf{0.110} & \textbf{0.059} & 0.079\\
         \bottomrule
    \end{tabular}
    }
    \label{ablation-2}
\end{table}

\section{Conclusion}
This paper proposes a multitemporal hyperspectral image unmixing method MUFormer. Different from previous methods, we propose Change Enhancement Module and Global Awareness Module to extract image information from three dimensions of time, space and spectra, and dynamically adapt to the changes of adjacent phases from the perspective of deep learning by using the advantages of transformer in processing long sequence information. The results of multiple datasets show that our model has achieved excellent results in multitemporal hyperspectral image endmember and abundance map estimation, and our method has also achieved a great advantage in running speed. We believe that the problem to be further solved for multitemporal hyperspectral image unmixing is the subtle endmember changes and abundance changes between different phases. At the same time, it is also a feasible new idea to integrate the denoising module into the unmixing, because we find that the multitemporal hyperspectral image is more susceptible to noise in the unmixing process. We hope that our model can shed new light on the task of multitemporal hyperspectral image unmixing.

\section*{Acknowledgments}
We would like to acknowledge Ricardo Borsoi for providing the dataset and its accompanying documentation.


\bibliography{content}

\newpage

\end{document}